\documentclass[twocolumn,showpacs,showkeys]{revtex4}
\usepackage{epsfig}
\usepackage{amsmath}

\begin{document}
\title{Probing phase transition order
of $q$-state Potts models using Wang-Landau Algorithm} 
\author{Tasrief  Surungan$^1$}
\email{tasrief@unhas.ac.id}
\author{Yukihiro Komura$^2$}
\email{y-komura@phys.se.tmu.ac.jp}
\author{Yutaka Okabe$^2$}
\email{okabe@phys.se.tmu.ac.jp}
\affiliation{$^1$Department of Physics, Hasanuddin University, Makassar, South Sulawesi 90245, Indonesia\\
$^2$Department of Physics, Tokyo Metropolitan University, Hachioji, 
Tokyo 192-0397, Japan }

\begin{abstract}
Phase transitions are  ubiquitous phenomena,  exemplified by the melting 
of ice and spontaneous magnetization of magnetic material.  In general, 
a phase transition is associated with a symmetry breaking of a system;  
occurs due to the competition between coupling interaction and external 
fields such as thermal energy.  If the phase transition occurs with no 
latent heat, the system experiences continuous transition, also known as 
second  order phase transition.  The ferromagnetic $q$-state Potts model 
with $r$ extra invisible states, introduced by Tamura, Tanaka, and 
Kawashima [Prog. Theor. Phys. 124, 381 (2010)], is studied by using the 
Wang-Landau method.  The  density of states difference (DOSD), 
$\ln g(E +\Delta E) - \ln g(E)$,  is used to investigate the order of 
the phase transition and examine the critical value of $r$  changing the 
second  to the first order transition. 
\end{abstract}

\maketitle

\section{Introduction}
The study of phase transitions has been one of the main subjects in physics 
since the introduction of the Ising model \cite{Ising} mainly aimed  to explain 
the phenomenon of spontaneous magntization in ferromagnet (FM). 
In general, a phase transition is associated with a symmetry breaking \cite{Landau};  
occurs due to the competition between coupling interaction 
and  thermal energy.  Systems are in high degree 
of symmetry at high temperature because all configurational spaces 
are accessible.  The decrease in temperature will reduce
thermal fluctuation and the system will stay in some favorable states.
If the phase transition occurs with no latent heat, the system experiences
continuous transition, also known as second  order phase transition,
which is a transition between the ordered and the disordered state.

The order of phase transition  of a system is frequently
not quite obvious and sometimes becomes a subject of debate.
In this paper, we focus on the phase transition of $q$-state Potts model.
It is a magnetic model applicable to many physical systems 
such as  polycrystalline material, simple fluids, percolation 
problems, etc.\cite{Potts} This model is a generalization of Ising 
model which is  known to exhibit second order phase transition for 
$d$-dimensional case, with $d \geq 2$. 

The order of phase transition of $q$-state Potts models varies with $q$ and 
the spatial dimensions  as well as the spin coupling interaction.  It is well 
known that the two-dimensional (2D) FM case of the model experiences second 
order phase transition for  $q \leq 4$, and first order for otherwise.
In 3D case, the model experiences first order transition for $q \geq 3$.  
The anti-ferromagnetic (AF) Potts model is considered to be more complex than the FM case as
its behaviour depends strongly on the microscopic lattice
structure. A systematic study of 3D AF case by Yamaguchi and Okabe 
reported that the phase transition is second order for $q=3$ and $4$. There
exists zero  temperature transition for $q=5$ and for $q=6$ no order found 
at any temperatures \cite{Yamaguchi}.

Recently, Tamura {\it et al.} studied $(q,r)$-state  Potts, which is 
a $q$-state Potts model with $r$ invisible 
redundant states\cite{Tamura}.  The invisible states of the model
affect  the entropy but
do not contribute to the internal energy.
Although this model is a straightforward extension of the standard
ferromagnetic Potts model,
due to the effect of invisible states, 
  a spontaneous $q$ fold symmetry breaking of
a first order phase transition can occur in 2D case for
 $q=2,3$, and $4$.
For each $q$, there is a critical value of $r$  which can
change the second order transition of the corresponding
standard $q$-state Potts model into a first order. Thus, 
it is of theoretical interest to determine this critical value of $r$.

Several methods are used for the analysis  
of  phase transition. An example of this is the usage of
 probability distribution function combined with the histogram reweighting 
 for  the analysis of first-order transition. For this type of transition,
the multicanonical Monte Carlo method  calculating the energy density of
states (DOS) was shown to be effective \cite{Berg}.
Quite recently, Komura and Okabe \cite{Komura} studied the difference
of  DOS, $\Delta \ln g(E) = \ln g(E+ \Delta E) - \ln g(E)$,
and examined the behavior of the first-order transition
with this quantity in connection with Maxwell's
equal area rule.
The paper elaborates the analysis of
the density of state difference (DOSD)  obtained from Wang-Landau
algorithm \cite{Wang} in determining the order of phase 
transition of the 2D $(q,r)$-state Potts model.  
The remaining part of the paper is organized
as follows: Section II describes the model and the method. 
The result  is  discussed in Section III. Section IV is devoted
to the summary and concluding remark.  

\section{Model and Simulation Method}

The $(q,r)$-state Potts model  
 is written with the following Hamiltonian 
\begin{equation}
H = - J \sum_{\langle ij \rangle}  \delta_{ S_i, S_j}  \sum_{\alpha}^q
\delta_{S_i, \alpha} \delta_{S_j, \alpha}
\end{equation}
For any spin configuration, the Potts spin $ S_i$ on site i-$th$ can take one
of $q+r$ states, with
$1 \leq S_i \leq q$ and $q+1 \leq S_i \leq q+r$ are respectively
referred to as colored states and invisible states. The standard
Potts models is recovered if $r=0$. 
Summation is performed 
over all the nearest-neighbor pairs of spins on a square lattice with
ferromagnetic interaction ($J > 0$) and with periodic boundary condition. 
If two neighboring spins take the same state, then the energy coupling is 
$-J$, otherwise it becomes zero. It is to be noticed that even if
two neighboring spins are equal, the energy remains zero if spins
are in invisible state, i.e.,  $q+1 \leq S_i = S_j \leq q+r$. 
Therefore, the  ground state energy is $-2NJ$  with $N$ is the number of spins.

\begin{figure}[b]
\includegraphics[width=2.34in]{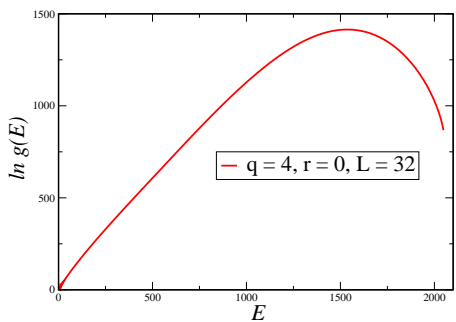}
\vspace{-0.1cm}
\caption{(Color online) The DOS plot of standard $q$-state Potts
model $(r = 0)$ with $q = 4$  for linear system
size $L=32$.}
\label{gmb01a}
\end{figure}

Monte Carlo simulation is a standard method
used in many field of physics, including  the
 study of phase transition of magnetic models.  
Here, we use Wang-Landau algorithm to calculate the DOS of the models.
This algorithm is particularly powerful for this purpose.
It has been implementend in various type of systems, such AF Potts model 
\cite{Yamaguchi} and fully frustrated Clock model \cite{Tasrief04}.
From the calculated  DOS $g(E)$, we can  straightforward
obtain DOSD defined as
\begin{equation}
\Delta \ln g = \ln g(E + \Delta E) - \ln g(E)
\end{equation}
For the considered model, $\Delta E = -J$, and the total energy
takes the value between $-2NJ$ and $0$, where $N$ is the number
of spins. However, for the convenience of the numerical
simulation, we shifted the energy so that it
takes the value between $0$ and $2NJ$. A general procedure of determining  
the order of
transition using DOSD  was described by Komura and Okabe \cite{Komura}. It was
shown that a system exhibiting  first-order transition will have
an S-like structure in the plot of $\Delta \ln g(E)$. 
The transition temperature of this type of transition
can be determined using  Maxwell's equal area rule.

\begin{figure}[b]
\includegraphics[width=2.4in]{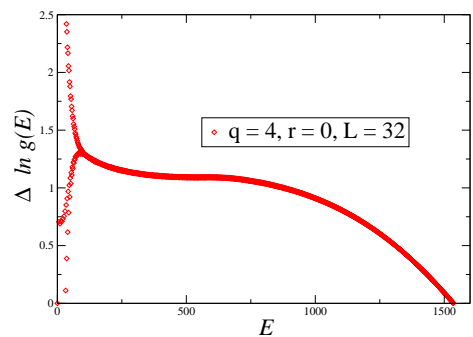}
\vspace{0.1cm}
\caption{(Color online) The DDOS plot of standard $q$-state Potts
model $(r = 0)$ with $q = 4$  for linear system
size $L=32$. Straight line is the inverse temperature
transition in the thermodynamic limit, $\beta_c = \ln(1 + \sqrt 4)$}
\label{gmb02a}
\vspace{-0.3cm}
\end{figure}

\section{Results and Discussion}

As has been described, the 2D $q$-state Potts
model undergoes  second-order phase transition for
$q \leq 4$ and  first order for otherwise. 
We should  show this behaviour using DOSD before presenting
the results for $(q,r)$-state Potts model.
The DOS of standard $q$-state Potts model for $q=3$ is plotted 
in Fig. \ref{gmb01a}. This type of figure is typical of DOS in
regular magnetic systems where density of ground state is 
much less than that of excited states. Obtaining the data
of any physical quantity  for every state, expressed 
as $Q(E)$ which can be referred as to density of quantity, 
we can calculate ensemble average $\langle Q \rangle$
via the following relation
\begin{equation}\label{QE}
\langle Q \rangle = \frac{\sum g(E) Q(E) e^{-E_i/kT}}{\sum g(E)e^{-E_i/kT}}
\end{equation}
This equation implies the forthcoming of Wang-Landau algorithm as
one can obtain the value of $\langle Q\rangle$ at any temperature.
In addition, from the same data of simulation
one can obtain by-product results for anti-ferromagnetic system,
namely by changing the energy $E_i$ into $-E_i$. This is demonstrated 
in the study of ferromagnetic $n$-state Clock model for various 2D lattices \cite{Tasrief13}.
From the data of DOS, shown in Fig. \ref{gmb01a}, 
we extract the DOSD of the corresponding model and system size, (See. Fig. \ref{gmb02a}).
Throughout, the energy is in unit of $J$ and
the Boltzmann constant is set as 1. 
\begin{figure}[t]
\includegraphics[width=2.5in]{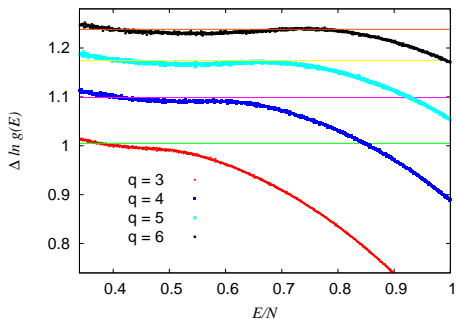}
\vspace{0.1cm}
\caption{(Color online) The DOSD plot of standard $q$-state Potts
model $(r = 0)$ with $q = 3, 4, 5$ and $6$  for linear system
size $L=32$. For the convenience, the
thermodynamic limit of  inverse temperature transition,
 $\beta_c = \ln (1+ \sqrt q)$, of each $q$ is also shown (dotted lines).}
\label{gmb03a}
\vspace{-0.2cm}
\end{figure}

The DOSD of standard  $q$-state Potts model $(r = 0)$ for
several values of $q$ ($q=3, 4, 5$ and $6$) 
are plotted  in Fig. \ref{gmb03a}. The linear system size $L$ is $32$. 
In order to reduce fluctuations,  in plotting
$\Delta \ln g(E)$ we use the data with the smoothing process,
$( \tilde f(E - 2 \Delta E) + 4 \star \tilde  f(E - \Delta E) 
 + 6 \star \tilde f(E) + 4 \star \tilde f(E +
\Delta E) + \tilde f(E + 2 \Delta E))/16 $  
with $\tilde f(E) = \Delta \ln g(E)$. 

\begin{figure}
\includegraphics[width=2.5in]{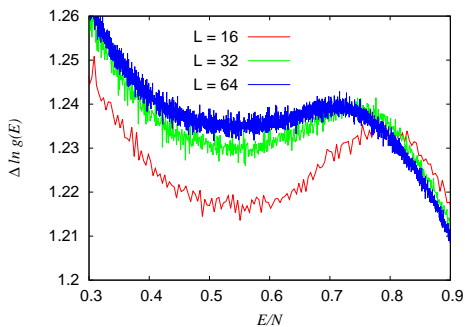}
\vspace{0.1cm}
\caption{(Color online) Size dependence of $\Delta \ln g(E)$ for $6$-state Potts model
($r = 0$) with 
linear system size L = 16,  32 and 64. The presence of S-like
structure is an essential indication of the first order transition.}
\label{gmb02}
\end{figure}

The S-like structure for $ q > 4 $ in Fig. \ref{gmb03a} is the indication of 
first order transition. The inverse transition temperature $\beta_c$ 
can be estimated by Maxwell's equal area rule as described in  Ref. \cite{Komura}. 
For convenience,  it is also shown $\beta_c$ in the thermodynamic limit
$\beta_c = \ln(1 + \sqrt q)$. We plot $\Delta \ln g(E)$ for 
the $6$-state Potts model for various sizes in Fig. \ref{gmb02}, which
shows size dependence; the S-like structure becomes smaller 
when the system size increases.  

Let us examine the meaning of $\Delta \ln g(E)$. This quantity
is related to the inverse of temperature in the microcanonical scheme
where we have 
\begin{equation}
\frac{\Delta \ln g(E)}{\Delta E} = \beta~ (\rm microcanonical)
\end{equation}
in the continuum limit. We note that $\Delta E/J = 1$ in the
present model. The plot of $ \Delta \ln g(E)$ as a function
 of $E$ as in Fig. \ref{gmb02} is nothing but $\beta$ versus $E$ plot; it is to be
noted that for discrete energy models such as the Potts
model there is an oscillating behavior for low DOS region
due to the discreteness as was shown in Fig. 1 of Ref. \cite{Komura}.
It is interesting to compare this with the temperature
dependence of the ensemble average of energy,  $\langle E \rangle$ following definition of
Eq. (\ref{QE})
\begin{figure}
\includegraphics[width=2.5in]{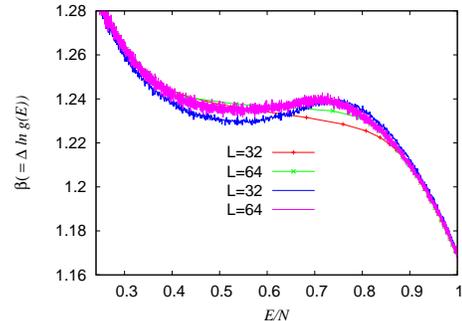}
\vspace{0.1cm}
\caption{(Color online) Comparison between plot of $\beta$ versus $E$ of
microcanonical and canonical schemes for $6$-state Potts
model. Solid curve indicates  plot of $\Delta \ln g(E)$ versus $E$,
whereas $\beta$ versus $\langle E(T)\rangle$ is
shown by dotted curve. The linear system size is L = 32 and
64.}
\label{gmb05a}
\end{figure}

In Fig. \ref{gmb05a}, we compare plot of $\Delta \ln g(E)$ versus $E$ and plot 
of $\beta$ versus $\langle E(T) \rangle$ for the $6$-state Potts model, with system
size $L = 32$ and $64$. The S-like structure shown by solid curve, 
in the microcanonical scheme leads
to the negative specific heat, or the thermodynamic instability.
Then, the canonical average of energy, which
is shown by dotted curve, has a jump at the first-order
transition temperature due to Maxwell's equal area rule.
When the system size increases from 32 to 64, the behavior
of S-like structure approaches that of the thermodynamic
limit; that is, the line to express the energy jump
becomes horizontal. Away from the first-order transition
temperature, two curves coincide with each other as the
system size increases. 

\begin{figure}
\includegraphics[width=2.5in]{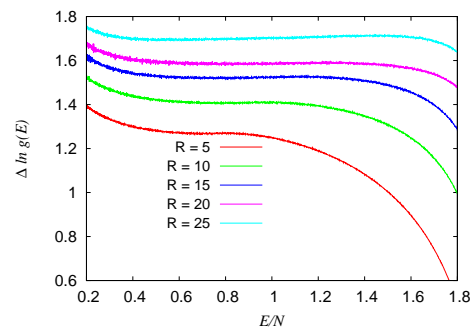}
\vspace{0.1cm}
\caption{(Color online) The DOSD plot of  $(4,r)$-state Potts model
with $r=5, 10, 20, 25$ and $30$ for system size $L=32$.}
\label{gmb06a}
\end{figure}

\begin{figure}
\includegraphics[width=2.5in]{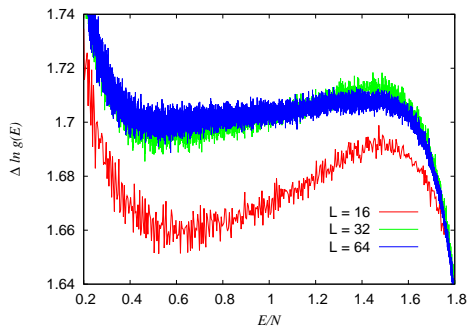}
\vspace{0.1cm}
\caption{(Color online) Size dependence of $\Delta \ln g(E)$ for
$q = 4$ and $r = 25$. The linear system size is $L = 16, 32$ and $64$. 
}
\label{gmb07a}
\end{figure}

We plot the DOSD for the $(4, r)$-state Potts
model of system size $L=32$, with $r = 5, 10, 15, 20$ and $25$ in Fig. \ref{gmb06a}. 
As indicated, there is an S-like structure for $r = 25$, whereas there is no such
a structure for $ r \leq 15$. The intermediate behavior is for
$r = 20$. To see the size dependence carefully, we plot the
size dependence of $\Delta \ln g(E)$ for $q = 4$ and $r = 25$, as a
typical example for the first-order transition, in Fig. \ref{gmb07a}.

We can estimate
$\beta_c$ and the interfacial free energy from the S-like curve
for each size. 
The procedure for estimating the transition temperature
and the interfacial free energy is the same as that for the
$q$-state Potts model \cite{Berg}. The numerically exact estimate
of the interfacial free energy for the q-state Potts model
was made by Borgs and Janke \cite{Jangke}. 
The detail analysis of obtaining these quantities for
$(q,r)$-state Potts model will be reported elsewhere.

\vspace{0.2cm}
\section{Summary and Concluding Remarks}\label{four}

In summary, we have studied the  $q$-state Potts model with and without 
invisible states and demonstrated the implementation of 
the density of states difference (DOSD) extracted from the main result of Wang-Landau 
Algorithm in examining the order of phase transition.  
We have obtained deeper understanding of the role of redundant
states in the $(q,r)$-state Potts model, namely transforming the
second order into first order phase transition at critical value of $r$.
For $q=4$ for example, the critical value for $r$ is
around 20.   There are several models where
the order of the transition changes with some parameter,
for example, the modified XY model \cite{Domany}. 
 It will be interesting to study the change of
  the order of transitions for these models
  using the present method of analyzing the density
  of states difference.

\section*{Acknowledgments}

The authors wish to thank Terry Mart,  Bansawang BJ and Safaruddin A. Prasad for valuable discussions.  
 The extensive computation was performed using the parallel computer facilities of
 the Department of Physics, Hasanuddin  University and that of
Tokyo Metropolitan University, JAPAN.
The present work is financially supported by Research Grant of BOPTN 
of Hasanuddin University, FY 2013.


\begin{thebibliography}{99}
\bibitem{Ising} E. Ising, Z. Phys., 31, 253, (1925).
\bibitem{Landau} L. D. Landau, {\it On the theory of phase transition}, 
in {\it Collected Papers of L. D. Landau}, edited by D. T. Haar (Pergamon Press, 1965).
\bibitem{Potts} R. B. Potts, Proc. Cambridge. Philos. Soc. 48, 106
(1952); F. Y. Wu, Rev. Mod. Phys. 54, 235 (1982).
\bibitem{Yamaguchi} C. Yamaguchi and Y.Okabe, J. Phys. A 34, 8781 (2001)
\bibitem{Tamura} R. Tamura, S. Tanaka, and N. Kawashima, Prog. Theor.
Phys. 124, 381 (2010).
\bibitem{Berg} B. A. Berg and T. Neuhaus, Phys. Rev. Lett. 68, 9
(1992).
\bibitem{Komura} Y. Komura and Y. Okabe, Phys. Rev. E 85, 010102(R) (2012).
\bibitem{Tasrief13}T. Surungan, {\it in preparation}.
\bibitem{Wang}F. Wang and D.P. Landau, Phys. Rev. Lett. 86, 2050
(2001); Phys. Rev. E 64, 056101 (2001).
\bibitem{Tasrief04} T. Surungan, Y. Okabe, and Y. Tomita,  J. Phys. A {\bf 37}, 4219 (2004).
\bibitem{Jangke} W. Janke, Phys. Rev. B 47, 14757 (1993);  C. Borgs and W. Janke, J. Phys. (France) I 2, 2011
(1992).
\bibitem{Domany} E. Domany, M.Schick, and R.H. Swendsen, Phys. Rev.  Lett. 52, 1535 (1984).
\end{thebibliography}
\end{document}